\documentclass[aps,prl,twocolumn,showpacs,amsmath,amssymb,amsfonts,nofootinbib,long]{revtex4}
\begin{document}
\title{Scaling Laws in Magnetohydrodynamic Turbulence}
\author{Leonardo Campanelli$^{1,2}$}
\email{campanelli@fe.infn.it}
\affiliation{$^{1}${\it Dipartimento di Fisica, Universit\`a di Ferrara, I-44100 Ferrara, Italy
\\           $^{2}$INFN - Sezione di Ferrara, I-44100 Ferrara, Italy}}
\date{October 4, 2004}


\begin{abstract}
We analyze the decay laws of the kinetic and magnetic energies and
the evolution of correlation lengths in freely decaying
incompressible magnetohydrodynamic (MHD) turbulence.

Scale invariance of MHD equations assures that, in the case of
constant dissipation parameters (i.e. kinematic viscosity and
resistivity) and null magnetic helicity, the kinetic and magnetic
energies decay in time as $E \sim t^{-1}$, and the correlation
lengths evolve as $\xi \sim t^{1/2}$.

In the helical case, assuming that the magnetic field evolves
towards a force-free state, we show that (in the limit of large
magnetic Reynolds number) the magnetic helicity remains constant,
the kinetic and magnetic energies decay as $E_v \sim t^{-1}$ and
$E_B \sim t^{-1/2}$ respectively, while both the kinetic and
magnetic correlation lengths grow as $\xi \sim t^{1/2}$.
\end{abstract}


\maketitle


Magnetic fields are observed in all gravitationally bound
large-scale structures in the Universe. They have been detected in
galaxies, in galaxy clusters, and there are strong hints that they
exist in superclusters, and in galaxies at high redshifts. These
last astronomical observations support the conjecture that
magnetic fields have been generated in the early Universe by
microphysics processes (for a full discussion see
Ref.~\cite{CoMF1,CoMF2,CoMF3,CoMF4,CoMF5,CoMF6,CoMF7} and
references therein). The study of the evolution of primordial
magnetic fields has been developed in the framework of the
so-called freely decaying magnetohydrodynamic (MHD) turbulence.
Numerical and analytical studies show that the relevant integral
quantities in MHD turbulence, as for example the magnetic energy
and correlation length, evolve in time following simple power
laws~\cite{Ole97,dMHD0,dMHD1,dMHD2,dMHD3,dMHD4,dMHD5,dMHD6,dMHD7,dMHD8}.
In this paper we shall give a possible explanation of those laws
in the light of some recent high-resolution numerical simulations
performed by Biskamp and M\"{u}ller~\cite{Bis99}, and Christensson
et al.~\cite{Chr02}.

We start by writing down the magnetohydrodynamic equations for a
incompressible fluid in the case in which the expansion of the
Universe can be neglected~\cite{Bis93}:
\begin{eqnarray}
\label{Eq1} & & \frac{\partial {\textbf v}}{\partial t} +
({\textbf v} \cdot \nabla){\textbf v} \, = \, - \nabla p \, +
{\textbf J} \times {\textbf B} + \nu \nabla^2 {\textbf v},
\\
\label{Eq2} & & \frac{\partial {\textbf B}}{\partial t} \, = \,
\nabla \times ({\textbf v} \times {\textbf B}) + \eta \nabla^2
{\textbf B},
\\
\label{Eq3} & & \nabla \cdot {\textbf v} \, = \, \nabla \cdot
{\textbf B} \, = \, 0,
\end{eqnarray}
where ${\textbf v}$ is the velocity of bulk fluid motion,
${\textbf B}$ is the magnetic field, ${\textbf J} = \nabla \times
{\textbf B}$ is the magnetic current, and $p$ is the thermal
pressure of the fluid
\footnote{The thermal pressure $p$ is not an independent variable;
indeed, taking the divergence of Eq.~(\ref{Eq1}) we can express
$p$ as a function of ${\textbf B}$ and ${\textbf v}$ as $\nabla^2
p = \nabla \cdot [{\textbf J} \times {\textbf B} - ({\textbf v}
\cdot \nabla){\textbf v}]$.}.
The kinematic viscosity $\nu$ and the resistivity $\eta$ are
dissipative parameters, and are determined by microscopic physics.
It is useful to define the kinematic and magnetic Reynolds
numbers, ${\text{Re}} = v l/\nu$ and ${\text{Re}}_B = vl/\eta$,
where $v$ and $l$ are the typical velocity and length scale of the
fluid motion. We say that the dynamics is turbulent when
${\text{Re}}$ and ${\text{Re}}_B$ are much greater than
unity~\cite{Ver04}.

In the case of the expanding Universe (with zero curvature and in
the radiation era), it has been shown that the MHD equations are
the same as the Eq.~(\ref{Eq1})-(\ref{Eq3}) provided that time,
coordinates and dynamical variables are replaced by the following
quantities~\cite{Bra96}:
\begin{eqnarray}
\label{Eq3'} & & \!\!\!\!\!\!\!\!\!\! t \rightarrow \tilde{t} =
\int \! a^{-1 }dt, \;\; {\textbf x} \rightarrow \tilde{{\textbf
x}} = a {\textbf x},
\\
\label{Eq3''}  & & \!\!\!\!\!\!\!\!\!\! {\textbf B} \rightarrow
\tilde{{\textbf B}} = a^2 {\textbf B}, \;\; \nu \rightarrow
\tilde{\nu} = a^{-1 } \nu, \;\; \eta \rightarrow \tilde{\eta} =
a^{-1 }\eta,
\end{eqnarray}
where $a$ is the expansion parameter, and we note that ${\textbf
v}$ is not scaled. Because of the formal coincidence of the MHD
equations in the expanding and non-expanding Universe, we can
study the evolution of the integral quantities in MHD turbulence
in both cases in a similar way. For definiteness, in this paper we
shall consider only the case of non-expanding Universe.

Let us introduce the magnetic energy density of an isotropic
plasma in a volume $V = \int_{2\pi/K}^L d^{\,3} x$ as
\begin{equation}
\label{Eq4} E_B(t) = \frac{1}{2V} \int_V \! d^{\,3} x \, {\textbf
B}^2({\textbf x},t) = \int_{2\pi/L}^{K} \!\!\! dk \, {\mathcal
E}_B(k,t),
\end{equation}
where ${\mathcal E}_B(k,t) = \frac{2\pi}{V} \, k^2 \, {\textbf
B}({\textbf k}) \! \cdot \! {\textbf B}^*({\textbf k})$ is the
magnetic energy density spectrum, ${\textbf B}({\textbf k})$ being
the magnetic field in Fourier space, and $k = |{\textbf k}|$.
Here, $2\pi/L$ and $K$ are the infrared and ultraviolet cutoffs,
respectively. In the following we shall assume that $L \rightarrow
\infty$ and $K \rightarrow \infty$.
\\
The expressions for the kinetic energy $E_v$ and the kinetic
energy spectrum ${\mathcal E}_v$ are similar to the magnetic ones,
with ${\textbf B}$ replaced by ${\textbf v}$.

The magnetic helicity density is
\begin{equation}
\label{Eq5} H_B(t) = \frac{1}{V} \int_V \! d^{\,3} x \, {\textbf
A} \cdot {\textbf B} = \int_{0}^{\infty} \!\!\! dk \, {\mathcal
H}_B(k,t),
\end{equation}
where ${\mathcal H}_B(k,t) = \frac{4\pi}{V} \, k^2 {\textbf
A}({\textbf k}) \! \cdot \! {\textbf B}^*({\textbf k})$ is the
magnetic helicity density spectrum
\footnote{The magnetic energy spectrum and magnetic helicity
spectrum are not independent, since any magnetic field
configuration satisfies the realizability condition $|{\mathcal
H}_B(k,t)| \leq 2k^{-1} {\mathcal E}_B(k,t)$ (see
e.g.~\cite{Fie00}). The field is said to be ``maximally helical"
if, for all $k$, ${\mathcal H}_B$ is of the same sign and
saturates the above inequality.},
and ${\textbf A}$ is the vector potential.

The relevant length scale in turbulence theory is the so-called
correlation length, which is the characteristic length associated
with the large magnetic energy eddies of turbulence. It is defined
by
\begin{equation}
\label{Eq6} \xi_B(t) = 2 \pi \, \frac{\int_{0}^{\infty} \! dk \,
k^{-1} {\mathcal E}_B(k,t)}{\int_{0}^{\infty} \! dk \, {\mathcal
E}_B(k,t)} \,.
\end{equation}
One can define a kinetic correlation length $\xi_v$ in the same
manner as the magnetic one, with ${\mathcal E}_B$ replaced by
${\mathcal E}_v$.

\section*{\normalsize Non-helical turbulence}

It is well known that the MHD equations, under the scaling
transformations ${\textbf x} \rightarrow \ell \, {\textbf x}$, $t
\rightarrow \ell^{1-h} \, t$, admit solutions of the type
\begin{eqnarray}
\label{Eq7} {\textbf v}(\ell \, {\textbf x},\ell^{1-h} \, t) \!\!&
= &\!\!
                   \ell^{\,h} \, {\textbf v}({\textbf x},t), \\
\label{Eq8} {\textbf B}(\ell \, {\textbf x},\ell^{1-h} \, t) \!\!&
= &\!\!
                   \ell^{\,h} \, {\textbf B}({\textbf x},t),
\end{eqnarray}
provided that the dissipative parameters $\nu$ and $\eta$ scale as
\begin{eqnarray}
\label{Eq9} \nu(\ell^{1-h} \, t) \!\!& = &\!\! \ell^{1+h} \, \nu(t), \\
\label{Eq10} \eta(\ell^{1-h} \, t) \!\!& = &\!\! \ell^{1+h} \,
\eta(t).
\end{eqnarray}
Here $\ell > 0$ is  the ``scaling factor'' and $h$ is a arbitrary
real parameter. Starting from the scaling relations
(\ref{Eq7})-(\ref{Eq10}), in the seminal paper~\cite{Ole97},
Olesen obtained the following expression for the magnetic energy
spectrum:
\begin{equation}
\label{Eq11} {\mathcal E}_B(k,t) = \lambda_B k^p \,
\psi_B(k^{\frac{3+p}{2}} t),
\end{equation}
where $\lambda_B$ is a constant, $\psi_B$ is a arbitrary
scaling-invariant function, and $p=-1-2h$. For a theory for which
$\nu$ and $\eta$ are constants, we must take $h=-1$, corresponding
to $p=1$. Numerical simulations show that the scaling invariance
of the solutions of MHD equations is approached only
asymptotically for $t \geq t_s$, where $t_s$ is un unknown
parameter. Hence, the scaling exponent $p$ is not fixed by initial
energy spectrum, the scaling law (\ref{Eq11}) being in general not
valid at $t=0$
\footnote{The fact that Eq.~(\ref{Eq11}) cannot be true at $t=0$
it easily understood. Assuming the validity of Eq.~(\ref{Eq11})
led to the inconsistence that the magnetic energy, $E_B(t) =
(\lambda_B/2t) \int_0^{\infty} \! dx \, \psi_B(x)$, diverges for
$t = 0$ (here, we have considered, for simplicity, the case
$p=1$).}.
Indeed, if we differentiate Eqs.~(\ref{Eq9}) and (\ref{Eq10}) with
respect to $\ell$, and put $\ell = 1$ afterwards, we get $\nu \sim
\eta \sim t^{\, (1-p)/(3+p)}$. We conclude that $p$ depends only
on the scaling properties of dissipation parameters. Now,
inserting Eq.~(\ref{Eq11}) in Eq.~(\ref{Eq4}) we have
\begin{equation}
\label{Eq12} E_B(t) = E_B(t_s) \left( \frac{t}{t_s} \right)^{\!
-\frac{2(1+p)}{3+p}} \!\!,
\end{equation}
where
\begin{equation}
\label{Eq12'} E_B(t_s) = \frac{2\lambda_B}{3+p} \,
t_s^{-\frac{2(1+p)}{(3+p)}} \! \int_0^{\infty} \!\! dx \,
x^{\frac{p-1}{p+3}} \, \psi_B(x),
\end{equation}
and it is assumed that, due to the presence of dissipation terms,
the integral is convergent. (Here and in the following we tacitly
assume that scaling laws are valid only for $t \geq t_s$). Because
${\textbf v}$ scales the same way as ${\textbf B}$,
Eqs.~(\ref{Eq11}) and (\ref{Eq12}) hold also in the case of
kinetic energy. In the case of constant dissipation parameters
(i.e. $p=1$) we get $E_B(t) \sim t^{-1}$.

Following the same procedure performed in Ref.~\cite{Ole97}, we
get for the magnetic helicity spectrum the expression: ${\mathcal
H}_B = \mu_B k^{p-1} \phi_B(k^{(3+p)/2} t)$, where $\mu_B$ is a
constant and $\phi_B$ is an arbitrary scaling-invariant function.
Integration with respect to $k$ gives $H_B(t) = H_B(t_s)
(t/t_s)^{-2p/(3+p)}$.
\\
Theoretical arguments (see e.g. \cite{Bis93,dMHD3}) and numerical
simulations of MHD equations (see e.g. \cite{Bis99,Chr02}) have
clearly show that the magnetic helicity is an approximately
conserved quantity in MHD turbulence (i.e. for large Reynolds
numbers $H_B \simeq {\mbox{const}}$).
In the light of this, we are led to the conclusion that all the
above scaling arguments can correctly describe the decay laws in
freely decaying MHD only in the case $H_B(t_s) = 0$, or in others
words, in the case of null magnetic helicity.

As regarding the correlation length, inserting Eq.~(\ref{Eq11})
into Eq.~(\ref{Eq6}) we obtain
\begin{equation}
\label{Eq13} \xi_B(t) = \xi_B(t_s) \left( \frac{t}{t_s}
\right)^{\! \frac{2}{3+p}} \!\!,
\end{equation}
where
\begin{equation}
\label{Eq13'} \xi_B(t_s) = 2\pi t_s^{\frac{2}{3+p}} \: \frac{
\int_{0}^{\infty} \! dx \, x^{\frac{p-3}{p+3}} \, \psi_B(x)}{
\int_{0}^{\infty} \! dx \, x^{\frac{p-1}{p+3}} \, \psi_B(x)} \, .
\end{equation}
In the case of constant dissipation parameters we get $\xi_B \sim
t^{1/2}$. The scaling law (\ref{Eq13}) is also valid for the case
of kinetic correlation length.

The evolution laws (\ref{Eq12}) and (\ref{Eq13}) for $p=1$ are
exactly the laws found by Biskamp and M\"{u}ller~\cite{Bis99} in
high-resolution numerical simulations of non-helical MHD
turbulence.

\section*{\normalsize Helical turbulence}

Mechanisms for generating helical magnetic fields in the early
Universe has been proposed during the last
years~\cite{Heli1,Heli2,Heli3,Heli4}. In the helical case, the
magnetic field is expected to evolve towards a substantially
force-free configuration, ${\textbf J} \times {\textbf B} = 0$.
This feature of helical MHD is expected on theoretical
ground~\cite{Tay74,Fie00}, and verified in numerical
simulations~\cite{Men81}.
\\
Assuming force-freedom [that is taking ${\textbf J} \times
{\textbf B} = 0$ in Eq.~(\ref{Eq1})], and assuming that $\nu$ and
$\eta$ scale as in Eqs.~(\ref{Eq9}) and (\ref{Eq10}), we find that
the helical MHD equations admit solutions of the type
\begin{eqnarray}
\label{Eq14} {\textbf v}(\ell \, {\textbf x},\ell^{1-h} \, t)
\!\!& = &\!\!
                          \ell^{\,h} \, {\textbf v}({\textbf x},t), \\
\label{Eq15} {\textbf B}(\ell \, {\textbf x},\ell^{1-h} \, t)
\!\!& = &\!\!
                           \ell^{\,m} \, {\textbf B}({\textbf x},t),
\end{eqnarray}
where $m$ is a new scaling exponents. Proceeding as in
Ref.~\cite{Ole97}, we obtain the scaling relations
\begin{eqnarray}
\label{Eq16} {\mathcal E}_B(k,t) \!\!& = & \!\! \lambda_B k^q \,
\psi_B(k^{\frac{3+p}{2}} t),  \\
\label{Eq17} {\mathcal E}_v(k,t) \!\!& = & \!\! \lambda_v k^p \,
\psi_v(k^{\frac{3+p}{2}} t), \\
\label{Eq18} {\mathcal H}_B(k,t) \!\!& = & \!\! \mu_B k^{q-1}
\phi_B(k^{\frac{3+p}{2}} t),
\end{eqnarray}
where $\lambda_B$, $\lambda_v$, $\mu_B$ are constants, $\psi_B$,
$\psi_v$, $\phi_B$ arbitrary scaling-invariant functions,
$p=-1-2h$, and $q=-1-2m$. Equations (\ref{Eq16}) and (\ref{Eq18})
imply the following scaling laws, respectively:
\begin{eqnarray}
\label{Eq19} E_B(t) \!\! & = & \!\! E_B(t_s) \left( \frac{t}{t_s}
\right)^{\! -\frac{2}{3+p} -2r} \!\!,  \\
\label{Eq20} H_B(t) \!\! & = & \!\! H_B(t_s) \left( \frac{t}{t_s}
\right)^{\! -2r} \!\!,
\end{eqnarray}
where we have defined $r = q/(3+p)$ for notational convenience,
and
\begin{eqnarray}
\label{Eq19'} \!\!\!\!\!\!\! E_B(t_s) \!\! & = & \!\!
\frac{2\lambda_B}{3+p} \, t_s^{-\frac{2}{3+p} -2r} \!\!
\int_0^{\infty} \!\! dx \, x^{\frac{2}{3+p}
+ 2r - 1} \psi_B(x) , \\
\label{Eq20'} \!\!\!\!\!\!\! H_B(t_s) \!\! & = & \!\!
\frac{2\mu_B}{3+p} \, t_s^{-2r} \!\! \int_0^{\infty} \!\! dx \,
x^{2r-1} \phi_B(x) .
\end{eqnarray}
The kinetic energy scales in time as in Eq.~(\ref{Eq12}), while
the magnetic and kinetic correlation lengths follow the same law
as in Eq.~(\ref{Eq13}). Since the magnetic helicity is a
quasi-conserved quantity in MHD turbulence, we expect that $r
\rightarrow 0$ for large Reynolds numbers. In order to find a
relation between these quantities, we shall apply the so-called
``Taylor variational principle''~\cite{Tay74,Bis93}. This
principle states that in magnetohydrodynamic turbulent systems,
the asymptotic state is expected to be a minimum-energy state
under the constraint $H_B = {\mbox{const}}$. This means that the
asymptotic state satisfies the variational equation $\delta [E -
\beta H_B] = 0$, where $E = E_v + E_B$ is the total energy, and
$\beta$ is a Lagrangian multiplier. Variation with respect to
${\textbf A}$ gives
\begin{equation}
\label{Eq21} {\mathbf J} = 2\beta {\mathbf B},
\end{equation}
while variation with respect to ${\textbf v}$ gives ${\textbf v} =
0$. Thus, the asymptotic state corresponds to a force-free
magnetic field configuration
\footnote{As shown by Field and Carroll~\cite{Fie00}, a maximally
helical magnetic field, whose energy spectrum is strongly peaked
at some wavenumber $k_p$, will be substantially force-free:
${\mathbf J} \simeq k_p {\mathbf B}$.}
with a vanishing kinetic energy
\footnote{We shall see that the kinetic energy decays faster then
the magnetic energy (for the case of constant dissipative
parameters and high Reynolds numbers we shall get $E_v \sim
t^{-1}$ and $E_B \sim t^{-1/2}$) and this means that the kinetic
energy is asymptotically negligible compared to the magnetic
energy.}.

It is worthwhile to stress that the force-free hypothesis
(\ref{Eq21}) is only valid in the large-scale range (i.e. small
$k$-scales). At small scales (large $k$-scales), i.e. in the
dissipative range, turbulence is not efficient, and then the
force-free hypothesis does not longer hold. Therefore, one has to
consider the presence of small-scale structures of the magnetic
field when investigating the dissipation of energy and helicity.
Theoretical arguments (see e.g. \cite{Bis93,dMHD3,Ole04}) seem to
indicate that the magnetic energy dissipation is finite at scale
$k_{\textmd{diff}} \sim \eta^{-1/2}$, while at this scale the
dissipation rate of the magnetic helicity goes to zero. We shall
see in the Appendix that this means that the presence of
small-scale structures of the magnetic field does not affect
significantly the decaying of helicity, while the dissipation of
energy is completely ruled by the decay of these small-scale
modes. Since in the following we shall analyze the large-scale
structure of the magnetic field, and in particular the decay of
the magnetic helicity, we shall work in the force-free
approximation.

Multiplying Eq.~(\ref{Eq21}) by ${\textbf A}$ and then integrating
in $d^{\, 3} x$, we get $\beta = E_B / H_B$. Now, using the
equation for the dissipation of the magnetic helicity (see e.g.
Ref.~\cite{Bis93}),
\begin{equation}
\label{Eq0-v2} \frac{dH_B}{dt} = -2\eta \! \int \! d^{\,3} x \,
{\textbf J} \cdot {\textbf B},
\end{equation}
and taking into account Eq.~(\ref{Eq21}), we obtain
\begin{equation}
\label{Eq0'-v2} \frac{dH_B}{dt} = - 8 \eta \frac{E_B^2}{H_B} \: .
\end{equation}
Inserting Eqs.~(\ref{Eq19}) and (\ref{Eq20}) into the above
equation we get $r = 4 \eta(t_s) \, t_s \beta^2 (t_s)$. In order
to give a physical meaning to the scaling exponent $r$, let us
introduce the magnetic Taylor micro-scale, the diffusion scale,
the magnetic Reynolds number evaluated using the magnetic Taylor
micro-scale, and the ``eddy turnover time'':
\begin{eqnarray}
\label{Eq22} \xi_T \!\! & = & \!\! 2 \pi \,
\frac{B_{\textmd{rms}}}{J_{\textmd{rms}}} \, , \;\;
\xi_{\textmd{diff}} = 2 \pi \sqrt{\eta t} \, , \\
\label{Eq22'} {\mbox{Re}}_B \!\! & = & \!\! \frac{v_{\textmd{rms}}
\, \xi_T}{\eta} \, , \;\; \tau_{\textmd{eddy}} =
\frac{\xi_v}{v_{\textmd{rms}}} \, ,
\end{eqnarray}
where $B_{\textmd{rms}}$, $J_{\textmd{rms}}$ and
$v_{\textmd{rms}}$ are the RMS magnetic field, magnetic current
and velocity field respectively. Taking into account
Eqs.~(\ref{Eq19})-(\ref{Eq21}), we get $\xi_T \sim
\xi_{\textmd{diff}} \sim t^{2/(3+p)}$, and $\beta = \pi/\xi_T$.
Moreover, observing that $v_{\textmd{rms}} \propto E_v^{1/2}$, and
using the scaling laws for $E_v$, $\xi_T$, $\eta$ and $\xi_v$, we
find that ${\mbox{Re}}_B$ is constant, while
$\tau_{\textmd{eddy}}$ scales as $t$. Finally, we can write $r$ as
\begin{equation}
\label{Eq23} r = \left( \frac{\xi_{\textmd{diff}}}{\xi_T}
\right)^{\! 2} = \: \frac{t_s}{\tau_{\textmd{eddy}}} \:
\frac{\xi_v}{\xi_T} \: \frac{(2 \pi)^2}{\mbox{Re}_B} \, ,
\end{equation}
where all quantities are evaluated at $t = t_s$.

For the case of constant dissipation terms we get $E_B \sim
t^{-1/2 - 2r}$, and $H_B \sim t^{-2r}$. These scaling laws are in
agreement with the laws found by Christensson et al. using
different scaling arguments, and are in good agreement with
numerical simulation performed by the same authors~\cite{Chr02}.
For very large magnetic Reynolds number and for initial condition
such that $t_s \, \xi_v / (\tau_{\textmd{eddy}} \, \xi_T)$ is of
order of unity (as in Ref.~\cite{Chr02,Bis99}), we have $E_v \sim
t^{-1}$, $E_B \sim t^{-1/2}$, and $H_B \simeq {\mbox{const}}$.
These are exactly the evolution laws that Biskamp and M\"{u}ller
have found by direct integration of MHD equations~\cite{Bis99}.

It is important to stress that in Ref.~\cite{Chr02} the initial
spectra are taken to be ${\mathcal E}_v(k,0) \propto k^2$ and
${\mathcal E}_B(k,0) \propto k^4$, while in Ref.~\cite{Bis99} are
${\mathcal E}_v(k,0) \propto k^2$ and ${\mathcal E}_B(k,0) \propto
k^2$. Then, the interesting fact is that different power-law
initial conditions led to the same scaling laws. This suggest that
the scaling exponents we found do not depend on the particular
choice of initial energy spectra, although the functional form of
the scaling functions $\psi_B$, $\psi_v$, $\phi_B$, and the
characteristic time $t_s$ are expected to depend on the initial
conditions.

In summary, we have studied the evolution laws of the relevant
integral quantities in freely decaying incompressible MHD
turbulence. Using scale invariance of MHD equations and force-free
hypothesis we have shown that, for constant dissipation parameters
and very large Reynolds numbers, the kinetic and magnetic energies
and correlation lengths evolve in time according to the following
laws:
\\
{\it i}) Non-helical case: $E_v \sim E_B \sim t^{-1}$, $\xi_v \sim
\xi_B \sim t^{1/2}$;
\\
{\it ii}) Helical case: $H_B \simeq {\mbox{const}}$, $E_v \sim
t^{-1}$, $E_B \sim t^{-1/2}$, $\xi_v \sim \xi_B \sim t^{1/2}$.

Finally, taking into account the results of different numerical
simulations of MHD equations (in particular see Ref.~\cite{Bis99}
and Ref.~\cite{Chr02}) we are led to suspect that this scaling
laws are {\it universal}, in the sense that they do not depend on
the initial conditions.


\begin{acknowledgments}
I would like to thank P. Cea for reading the manuscript and for
helpful and stimulating discussions. I also thank A. Marrone, P.
Olesen, and A. Palazzo for very useful comments.
\end{acknowledgments}


\section*{\normalsize Appendix}

We start by noting that Eq.~(\ref{Eq19}) together with
Eq.~(\ref{Eq23}) implies that the decay of the magnetic energy is
ruled by a dissipative term:
\begin{equation}
\label{Eq1-v2} \frac{dE_B}{dt} = - 2 \eta \, k_T^2 E_B -
\frac{2}{3+p} \, \eta \, k_{\textmd{diff}}^2 E_B,
\end{equation}
where $k_T = 2\pi/\xi_T$, and $k_{\textmd{diff}} = 2 \pi/
\xi_{\textmd{diff}}$ is the $k$-scale at which the diffusion is
efficient. On the other hand, taking into account the evolution
equation for the magnetic energy (see e.g. Ref.~\cite{Bis93}),
\begin{equation}
\label{Eq2-v2} \frac{dE_B}{dt} = \: - \! \int \! d^3x \,
\textbf{v} \cdot \textbf{J} \times \textbf{B} \: - \: \eta \! \int
\! d^3x \, \textbf{J}^2,
\end{equation}
we get that, in force-free hypothesis, the decaying law of the
magnetic energy becomes $dE_B/dt = - 2 \eta k_T^2 E_B$, which
gives $E_B \sim t^{-2r}$. This means that the use of the force
free-hypothesis correctly reproduce the diffusive part of the
power in Eq.~(23), but cannot give the other part of the power,
$-2/(3+p)$, which must come from the small-scale modes of
$\textbf{B}$. In order to get the right decaying law for the
magnetic energy, we have to take into account the small-scale
modes of the magnetic field in the expression of the current. So
we write:
\begin{equation}
\label{Eq3-v2} \textbf{J} \simeq k_T(t) \textbf{B} +
k_{\textmd{diff}}(t) \textbf{b},
\end{equation}
where $\textbf{b}(\textbf{x})$ is the magnetic field on
dissipation scale. (It should be noted that in this case the
magnetic field is no longer force-free.) Because $\textbf{b}$ is a
tangled field defined only in the dissipative range, we expect
that $ \int \! d^3x \, \textbf{C} \, \cdot \, \textbf{b} \simeq
0$, where $\textbf{C}(\textbf{x})$ is an independent function of
$\textbf{b}(\textbf{x})$ (i.e. $\textbf{C}$ is any combination of
$\textbf{B}$ and $\textbf{v}$ only). It is straightforward to show
that the presence of small-scale structures of the magnetic field
does not affect the decaying of helicity, while the dissipation of
energy is completely ruled by the decay of these small-scale
modes. Indeed, inserting Eq.~(\ref{Eq3-v2}) into
Eq.~(\ref{Eq0-v2}) and Eq.~(\ref{Eq2-v2}) we get, respectively:
\begin{equation}
\label{Eq4'-v2} \frac{dH_B}{dt} \simeq - 4 \eta \, k_T E_B,
\end{equation}
(we remember that $k_T = 2\beta = 2 E_B/H_B$), and
\begin{equation}
\label{Eq4-v2} \frac{dE_B}{dt} \simeq - 2 \eta \, k_T^2 E_B -2
\eta \, k_{\textmd{diff}}^2 E_b.
\end{equation}
Comparing Eq.~(\ref{Eq1-v2}) and (\ref{Eq4-v2}), we conclude that
the presence of small-scale modes of $\textbf{B}$ explains the
$-2/(3+p)$ part of the power in Eq.~(\ref{Eq19}) provided that
\begin{equation}
\label{Eq5-v2} E_b \simeq \frac{2}{3+p} \, E_B.
\end{equation}
The interesting and perhaps unexpected fact is that the energy
associated to the small-scale modes, $E_b$, is comparable to the
total energy, $E_B$, of the magnetic field.


\end{document}